# Stability and Throughput of Buffered Aloha with Backoff

Tony T. Lee, *Fellow, IEEE*, and Lin Dai, *Member, IEEE*

*Abstract*—This paper studies the buffered Aloha with *K*-exponential backoff collision resolution algorithms. The buffered Aloha network is modeled as a multi-queue single-server system. We adopt a widely used approach in packet switching systems to decompose the multi-queue system into independent first-in-first-out (FIFO) queues, which are hinged together by the probability of success of head-of-line (HOL) packets. A unified method is devised to tackle the stability and throughput problems of *K*-exponential backoff with any cutoff phase *K*. For networks with a finite number of nodes, we show that the *K*-exponential backoff is stable if the retransmission factor is properly chosen from the stable region. The maximum stable throughput is derived and demonstrated via examples of geometric retransmission (*K*=1) and exponential backoff (*K*=∞). For networks with an infinite number of nodes, we show that geometric retransmission is unstable, and the stable network throughput of exponential backoff can only be achieved at the cost of potential unbounded delay in each input queue. Furthermore, we address the stability issue of the systems at the undesired stable point. All analytical results presented in this paper are verified and confirmed by simulations.

*Index Terms*—Random access, slotted Aloha, exponential backoff, geometric retransmission, stability

## I. INTRODUCTION

A fundamental problem of multi-access communications is how to efficiently share the channel resource among multiple users. From the Aloha network to today's IEEE 802.11 Wi-Fi network, random access has proven to be a simple yet elegant solution; transmit if there is a request, and back off if a collision occurs. The minimum coordination and distributed control have enabled random access to become one of the most widely deployed network technologies used today [1-3].

Throughput analysis on random access networks can be traced back to Abramson's landmark paper [1]. Assuming an infinite number of nodes, Abramson proposed to model the aggregate traffic as a Poisson random variable with parameter *G*, which captures the essence of contentions among users without intricate analysis. The steady-state equilibrium throughput *G*exp(-*G*) derived from this simplified model sheds useful insight into many aspects of network performance, such as the maximum network throughput of $e^{-1}$ when *G*=1.

Research interests have bifurcated since then. On one hand, intense activities have focused on the stability analysis of Abramson's Aloha protocol, under the original assumption that the network is saturated and each node always has a packet to transmit, or an infinite number of bufferless nodes. Most of these studies ignored queuing aspects, for instance, the offered load of each node, but emphasized the throughput of the whole network.

Specifically, early work showed that slotted Aloha is unstable with an infinite population [4-5]. To stabilize a finite-node Aloha system, a drift approach was developed to analyze the system transition so that the retransmission probability could be adjusted accordingly. This requires the knowledge of the number of backlogged nodes [6-9]. Rather than investigating how to estimate the backlogs [10-11], binary exponential backoff (BEB) algorithms were proposed to achieve stability by reducing the retransmission probability according to the number of collisions the packet has experienced [12]. It was proven in [13] that BEB is also unstable with an infinite number of nodes. Later, under a finite-node model, it was shown that BEB can be stable if the aggregate arrival rate is sufficiently small [14]. Different upper bounds of the aggregate arrival rate have been developed since then [15-16] but none of them has become the consensus [17]. The stability issue of random access protocols remains an open problem.

In the meantime, a great deal of effort has been made to establish a buffered Aloha model, which was initiated in [18] and further developed in [19-22]. With the interactions among the different queues taken into consideration, an *n*-node buffered Aloha system was modeled as an *n*-dimensional random walk and the exact rate region for the two-node case was derived in [18]. Unfortunately, the generalization of this approach to an arbitrary *n*-node system encountered tremendous difficulties. When *n* exceeds three, the interactions among the queues become unmanageable and the problem is intractable even for the simplest version of slotted Aloha [23], let alone an Aloha system with exponential backoff.

In this paper, the slotted Aloha is modeled as a multi-queue single-server system; each node is equipped with an infinite buffer and treated separately. A widely used approach in packet switching systems [24-26] is adopted to decompose the multi-queue system into independent FIFO queues with Bernoulli arrivals of rate $\lambda$ packets per time slot. These decomposed Geo/G/1 queues are then hinged together by the service time that is determined by the probability of success *p* of HOL packets.

The stability of networks with *n* nodes can be defined in terms of either the network throughput or delay. However, the two definitions are not exactly the same. Under the throughput definition, a network is stable if the network throughput $\hat{\lambda}_{out}$ equals the aggregate input rate $\hat{\lambda} = n\lambda$; or the departure rate equals the input rate. Under the delay definition, a network is stable if the offered load $\rho$ of each input Geo/G/1 queue is strictly less than 1. It is obvious that the delay stability implies the throughput stability, but the reverse is not necessarily true. A network with stable throughput that fails to ensure the delay bound is called *pseudo-stable*.

The probability of success of a buffered Aloha with *K*-exponential backoff has one desired stable point at $p_L$, one unstable equilibrium point at $p_S$ and one undesired stable point at $p_A$. Both $p_L$ and $p_S$ are roots of the characteristic equation of probability of success *p*, which is completely determined by the aggregate input rate $\hat{\lambda}$. The undesired stable point $p_A$, however, is dependent on the backoff parameters such as the retransmission factor *q* and the cutoff phase *K*.

Our analysis is concentrated on the specifications of the stable regions of retransmission factor *q*. In the absolute stable region, the network is guaranteed to converge to the desired stable point $p_L$ that ensures both throughput and delay stabilities. In a region with weaker assurance, called *asymptotic stable region*, the stability of exponential backoff can be ensured with a very high probability. Outside these desired stable regions, we show that the network with exponential backoff may poise on the undesired stable point $p_A$ with stable throughput but unbounded delay. A pseudo-stable region is introduced to manifest the resilience of exponential backoff in the face of transient fluctuations of input traffic. The stable regions and the associated maximum throughput in these regions of *K*-exponential backoff are outlined as follows. In particular, *geometric retransmission* (*K*=1) and *exponential backoff* (*K*=∞) are two stereotypes of interest.

**Geometric Retransmission:** For networks with a finite number of nodes *n*, the absolute stable region is given by:

$$\frac{\hat{\lambda}(1-p_L)}{p_L(n-\hat{\lambda})} \leq q \leq -\frac{\ln p_S}{n}$$

and the maximum stable throughput is $e^{-1}$. The network is unstable if the number of nodes *n* is infinite, which agrees with previous studies [4-5] that the slotted Aloha is inherently unstable as *n*→∞.

**Exponential backoff:** For networks with a finite number of nodes *n*, the absolute stable region is given by:

$$\frac{1-p_L}{1-\hat{\lambda}/n} \leq q \leq -\frac{\ln p_S}{n}$$

and the maximum stable throughput is ln*n*/*n*. It has been shown in [14-16] that the network with binary exponential backoff (BEB) is stable if the arrival rate is sufficiently small. Our result specifically shows that if the aggregate input rate $\hat{\lambda}$ is lower than $\frac{1}{2}ne^{-n/2}$, then the retransmission factor *q*=1/2 is inside the absolute stable region.

With an infinite number of nodes *n*, the absolute stable region becomes empty. Furthermore, we show that in the following asymptotic stable region

$$\frac{1-p_L}{1-\hat{\lambda}/n} \leq q \leq 1 - p_L - \frac{\ln p_S}{n}p_L,$$

the maximum throughput $e^{-1}$ is achievable with probability 1-ε, and ε→0 as the number of nodes *n*→∞. If the number of nodes *n* is infinite, the asymptotic stable region shrinks to a single point *q*=1-$p_L$, which coincides with the result claimed in [17]. However, the mean delay at this point would be unbounded and the network is pseudo-stable.

Below the unstable equilibrium point $p_S$, the probability of success will drift to the undesired stable point $p_A$. We prove that this undesired stable point $p_A$ converges to 1-*q* as *n*→∞, and the pseudo-stable region is given by:

$$1 - p_L \leq q \leq 1 - p_S.$$

In this region, the network is stable in terms of throughput, $\hat{\lambda}_{out} = \hat{\lambda}$, while at the cost of unbounded mean delay. Because of this pseudo-stability, exponential backoff is more adaptive to transient traffic fluctuations when compared to geometric retransmission. The network is unstable outside this pseudo-stable region as the throughput $\hat{\lambda}_{out} < \hat{\lambda}$.

The remainder of this paper is organized as follows. Section II establishes the queuing model and presents the preliminary analysis of buffered Aloha with backoff scheduling. Section III describes the absolute stable region, inside which the system can be guaranteed to converge to the desired stable point. Section IV introduces the asymptotic stability of exponential backoff. The analysis on the undesired stable point and the pseudo-stability of exponential are presented in Section V. The simulation results are provided in Section VI and conclusions are summarized in Section VII.

The main notations used in this paper are listed as follows:
*n*: number of nodes
*K*: cutoff phase
*q*: retransmission factor (0<*q*<1)
*λ*: input rate per node
$\hat{\lambda}$: aggregate input rate, also the steady-state network throughput. $\hat{\lambda}$=*nλ*
*ρ*: offered load of each node's queue
*p*: probability of success
*G*: attempt rate
$p_L$: desired stable point. $p_L = \exp\{W_0(-\hat{\lambda})\}$
$p_S$: unstable equilibrium. $p_S = \exp\{W_{-1}(-\hat{\lambda})\}$
$p_A$: undesired stable point
$q_l$: lower bound of retransmission factor *q*
$q_u$: upper bound of retransmission factor *q*
*S*: stable region of retransmission factor *q*. *S* =[$q_l$, $q_u$]
$\hat{\lambda}_{max\_S}$: maximum stable throughput. $\hat{\lambda}_{max\_S} = \sup_S \hat{\lambda}$

## II. PRELIMINARY ANALYSIS

The slotted Aloha resembles the statistical multiplexer in packet switching systems, both of which can be considered as a system with multiple input queues contending for a single server. The main difference lies in their contention resolutions. When there is more than one packet request for the output in the statistical multiplexer, one packet will be selected randomly and dispatched to the output channel. In the slotted Aloha, however, all packets contending for the same time slot will be dismissed.

In contrast to the classical infinite-node-no-buffering model [27], we will consider the slotted Aloha as a multi-queue single-server system, as shown in Fig. 1, in which each node is equipped with an infinite buffer and served by a single channel.

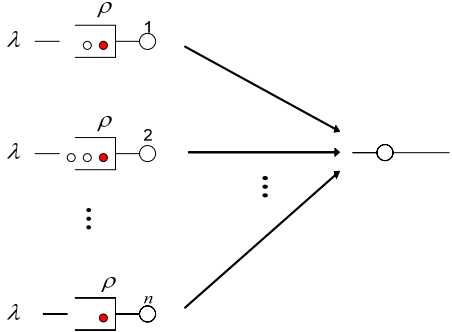

Fig. 1. Multi-queue single-server system

This section is devoted to the analysis of buffered Aloha based on the decomposition of the coupled nodes into independent Geo/G/1 queues, an approach similar to the widely studied model of packet switching systems [24-26]. These decomposed queues will then be glued together by the probability of success $p$ and the retransmission factor $q$ of the backoff protocols. Both analytical and simulation results in packet switching systems show that this approximation is an effective approach with high accuracy to model the multi-queue systems [26].

### A. Buffered Aloha with Backoff

The multi-queue single-server system with $n$ input queues can be characterized as a discrete time Markov process with a state space represented by the vector $(C_1, C_2, \ldots, C_n)$, where $C_i$ is the queue length of node $i$. This multi-dimensional Markov chain is obviously intractable as the number of nodes $n$ becomes too large. Thus, we adopt the decomposition approach in packet switching systems and regard each node as an independent FIFO queue with identical Bernoulli arrival processes of rate $\lambda$.

The contentions are resolved by backoff rescheduling of HOL packets. The number of collisions experienced by an HOL packet is called the *phase* of the packet. Initially, a fresh HOL packet is in phase 0, and it moves to the next phase if it is involved in a collision. Let $p$ be the probability of a successful transmission. A *K-exponential backoff* protocol allows a packet in phase $i$ to be transmitted with probability $q^i$, $i=0,1,\ldots,K$, where $q$ is the *retransmission factor* and $K$ is the *cutoff phase* of the protocol. The phase transition process of an HOL packet can be described by the Markov chain shown in Fig. 2.

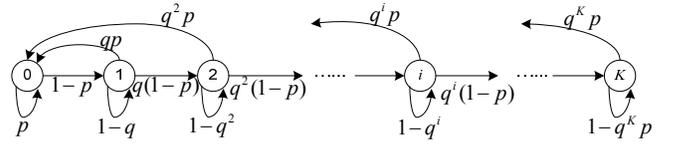

Fig. 2. State transition diagram of HOL packets.

Note that the probability of success $p$ in the Markov chain is assumed to be independent of the phase of the HOL packet. Intuitively, the chance that an HOL packet has a successful transmission should not vary with the number of collisions it has suffered. This assumption has been accepted and verified in various references [28].

Two particular backoff schemes are of special interest. *Geometric retransmission* is a special case with cutoff phase $K=1$, that is, the retransmission probability is a constant $q$ regardless of the number of collisions suffered. This collision resolution algorithm is the original version of slotted Aloha protocol that has been extensively investigated in [1, 4-11, 18-23]. If the cutoff phase is unlimited, $K=\infty$, then the protocol is simply called *exponential backoff*. For example, the binary exponential backoff (BEB) in previous studies [13-16] assumes $q=1/2$ and $K=\infty$.

Let $f_0, f_1, \ldots, f_K$ represent the limiting probabilities of the Markov chain shown in Fig. 2. We have

$$f_i = f_{i-1} q^{i-1}(1-p) + f_i(1-q^i), \ i=1,\ldots, K-1, \quad (1)$$

and

$$f_K = f_{K-1} q^{K-1}(1-p) + f_K(1-q^K p) \quad (2)$$

It follows from (1-2) that

$$f_0 = 1 / \left( \frac{q}{p+q-1} - \left( \frac{q}{p+q-1} - \frac{1}{p} \right) \cdot \left( \frac{1-p}{q} \right)^K \right), \quad (3)$$

$$f_i = f_0 \left( \frac{1-p}{q} \right)^i, \ i=1,\ldots, K-1, \ \text{and} \quad (4)$$

$$f_K = f_0 \left( \frac{1-p}{q} \right)^K / p. \quad (5)$$

**Lemma 1.** *For buffered Aloha with K-exponential backoff ($1 \leq K \leq \infty$), the offered load of each queue is $\rho = \lambda / f_0$.*

Proof: Since each fresh HOL packet will be in phase 0 for one time slot only, it will be either transmitted or blocked. Thus, the probability of finding a fresh HOL packet in a node buffer at any time slot, $\rho f_0$, should be equal to the input rate $\lambda$ of the Bernoulli arrival process. The offered load is therefore given by $\rho = \lambda / f_0$, and the mean service time of each HOL packet is $1/f_0$. □

The offered load $\rho$ per node depends on the probability of success $p$ and retransmission factor $q$, and it can be expressed as

$$\rho = \lambda / f_0 = \lambda \left( \frac{q}{p+q-1} - \left( \frac{q}{p+q-1} - \frac{1}{p} \right) \cdot \left( \frac{1-p}{q} \right)^K \right). \quad (6)$$

For geometric retransmission ($K=1$), the offered load $\rho$ is

$$\rho^{Geo} = \lambda(1-p+pq)/(pq). \quad (7)$$

In the case of exponential backoff ($K=\infty$), we have

$$\rho^{Exp} = \lambda q /(p+q-1). \quad (8)$$

Since the probability of success $p$ is determined by the activities of all nodes in the entire network, the offered load $\rho$ will serve as the glue to stick separately treated queues together in our analysis. In the original slotted Aloha model proposed by Abramson [1], the network throughput $\hat{\lambda} = G\exp(-G)$ was obtained by assuming no buffer for queuing at each node and considering the total number of attempts as a Poisson random variable with parameter $G$. The probability of success $p$ can then be expressed as $p = \hat{\lambda}/G = \exp(-\hat{\lambda}/p)$. In the next theorem, we will show that this characteristic of slotted Aloha still holds with an infinite buffer for queuing at each node, and it is invariant with respect to the retransmission factor $q$ and the cutoff phase $K$ under the independence assumption.

**Theorem 1.** *For buffered Aloha with K-exponential backoff ($1 \leq K \leq \infty$), the probability of success $p$ is given by*
$$p = \exp(-\hat{\lambda}/p), \quad (9)$$
*where $\hat{\lambda} = n\lambda$ is the aggregate input rate.*

Proof: Each node in the network must be in one of the following states:
State 1: idle;
State 2: busy with a fresh HOL;
State 3: phase $i$ and retransmitting, $i=1,2,\ldots,K$;
State 4: phase $i$ and not retransmitting, $i=1,2,\ldots,K$.

We know that the probability of a node being busy is $\rho$, and the probability of an HOL packet being in phase $i$ given that the node is busy is $f_i$, $i=0,1,\ldots,K$. Therefore, the probability of the above four states are given by:
1) Pr{node is in State 1}$=1-\rho$;
2) Pr{node is in State 2}$=\rho f_0$;
3) Pr{node is in State 3}$=\rho f_i q^i$, $i=1,2,\ldots K$;
4) Pr{node is in State 4}$=\rho f_i(1-q^i)$, $i=1,2,\ldots K$.

When a node successfully transmits a packet, its $n-1$ interfering nodes must be either in State 1 or State 4. Suppose that the number of idle nodes (State 1) and nodes in phase $i$ without retransmitting (State 4) are given by $m_{idle}$ and $m_i$, $i=1,\ldots,K$, respectively. We have $m_{idle} + \sum_{i=1}^{K} m_i = n-1$. The probability of success $p$ in steady-state conditions can then be written as

$$p = \sum_{\{m_{idle},m_1,\ldots,m_K\}} \frac{(n-1)!}{m_{idle}!m_1!\cdots m_K!} \Pr\{\text{node is in State 1}\}^{m_{idle}}$$
$$\cdot \prod_{i=1}^{K} \Pr\{\text{node is in State 4, phase } i\}^{m_i}$$
$$= (\Pr\{\text{node is in State 1}\} + \sum_{i=1}^{K} \Pr\{\text{node is in State 4, phase } i\})^{n-1}$$
$$= \left(1 - \rho + \sum_{i=1}^{K} \rho f_i(1-q^i)\right)^{n-1} \quad (10)$$

Substituting (3-6) into (10), we have
$$p = \left(1 - \lambda \sum_{i=1}^{K-1}(1-p)^i - \lambda\frac{(1-p)^K}{p}\right)^{n-1}$$
$$= (1-\lambda/p)^{n-1} \overset{\text{with a large } n}{\approx} \exp(-\hat{\lambda}/p). \quad \square$$

According to the above theorem, the expected number of attempts per time slot is given by
$$G = \hat{\lambda}/p = -\ln p. \quad (11)$$

It follows that the aggregate steady-state network throughput satisfies
$$\hat{\lambda} = G\exp(-G), \quad (12)$$
which peaks at $\hat{\lambda}_{max} = e^{-1}$ when the attempt rate $G=1$.

In spite of the fact revealed in Theorem 1 that the probability of success $p$ and the attempt rate $G$ are independent of the retransmission factor $q$ and the cutoff phase $K$, we will show that the selection of these backoff parameters, $q$ and $K$, is not arbitrary, and they are the keys to guarantee that the stable network throughput $\hat{\lambda}$ can be achieved by the backoff protocols.

*B. Stable Points of the Lambert W Function*

The solutions of the fundamental characteristic equation (9) of probability of success $p$, or its equivalent equation (12) of attempt rate $G$, can be represented by the Lambert W function defined by
$$W(z)e^{W(z)} = z, \quad (13)$$
which was first considered by J. Lambert around 1758, and later, studied by L. Euler [29].

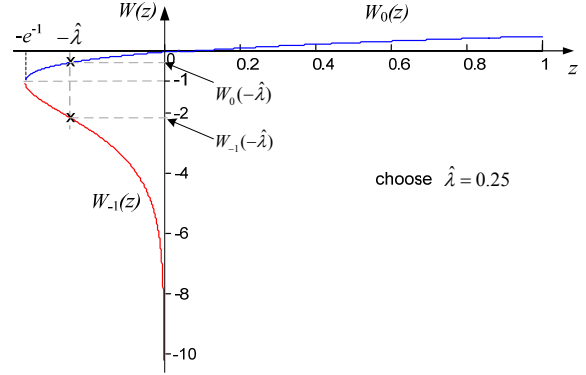

Fig. 3. The Lambert W function

The Lambert W function is a multivalued function. If $z$ is real and $-e^{-1} < z < 0$, there are two possible real values of $W(z)$: the principal branch $W_0(z) \in [-1,\infty]$ and the other branch $W_{-1}(z) \in [-\infty,-1]$. Both branches are illustrated in Fig. 3. The two non-zero solutions of (9) correspond, respectively, to the two branches of the Lambert W function, whose series expressions are given as follows:

1) $p_L = \exp(W_0(-\hat{\lambda}))$, and $W_0(z)$ has the following series expansion
$$W_0(z) = \sum_{i=1}^{\infty} \frac{(-i)^{i-1}}{i!} z^i = z - z^2 + \tfrac{3}{2}z^3 - \tfrac{8}{3}z^4 + \tfrac{125}{24}z^5 - \cdots \quad (14)$$
which can be derived by using the Lagrange inversion theorem. According to (14), $p_L$ can be further written as
$$p_L = \exp(W_0(-\hat{\lambda})) = 1 - \hat{\lambda} - \tfrac{1}{2}\hat{\lambda}^2 - \tfrac{2}{3}\hat{\lambda}^3 - \tfrac{63}{24}\hat{\lambda}^4 - \cdots \quad (15)$$
Usually $1-\hat{\lambda}$ is a good approximation for $p_L$ as $\hat{\lambda} \leq e^{-1}$.

2) $p_S = \exp(W_{-1}(-\hat{\lambda}))$, and $W_{-1}(z)$ has the following series expansion
$$W_{-1}(z) = \sum_{i=0}^{\infty} \mu_i x^i = -1 + x - \tfrac{1}{3}x^2 + \tfrac{11}{72}x^3 - \tfrac{43}{540}x^4 + \tfrac{769}{17280}x^5 - \cdots \quad (16)$$
in which $x = -\sqrt{2(ez+1)}$, and the coefficient $\mu_i$ is given in [29].

A good approximation for $p_S$ when $\hat{\lambda} > e^{-1}/2$ is given by
$$p_S = \exp(W_{-1}(-\hat{\lambda})) \approx \exp\{-\tfrac{5}{3} - \sqrt{2} + e(\tfrac{2}{3} + \tfrac{\sqrt{2}}{2})\hat{\lambda}\}. \quad (17)$$

The two non-zero roots $p_S = \exp(W_{-1}(-\hat{\lambda}))$ and $p_L = \exp(W_0(-\hat{\lambda}))$ are displayed in Fig. 4. As we can see, the roots of (9) are intersections of $y=p$ and $y = \exp(-\hat{\lambda}/p)$. In general, we have $p_S \leq p_L$ and the equality holds for $p_S = p_L = e^{-1}$ when $\hat{\lambda} = e^{-1}$.

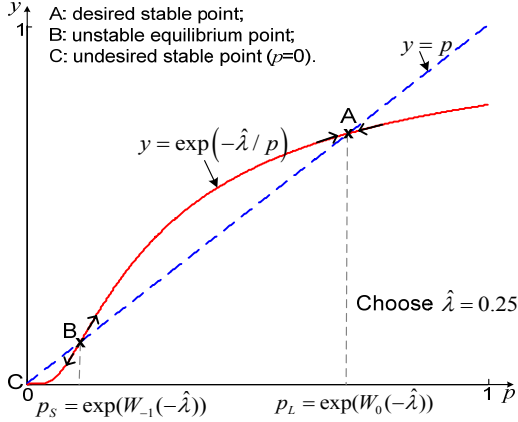

Fig. 4. Stable points of probability of success $p$

The previous stability analysis based on the drifts of the number of backlogged nodes revealed that the system has a desired stable point, an unstable equilibrium and an undesired stable point [6-7, 27]. The probability of success illustrated in Fig. 4 shows a desired stable point at $p_L$ and an undesired stable point at zero, while the point $p_S$ is an unstable equilibrium point. The dynamics of these stable points characterized in the following theorem confirms the observations made by the drift analysis.

**Theorem 2**: *Let $p_t$ be the probability of success at time slot t. If $p_t > p_S$ then $p_t \to p_L$ as $t \to \infty$.*

Proof: According to Theorem 1, the probability of success $p_{t+1}$ at time slot $t+1$ is determined by the loading $\rho_t$ of each queue in time slot $t$ as $p_{t+1} = \exp(-\hat{\lambda}/p_t)$. Given $p_t > p_S$, the probability of success $p_t$ will converge to the desired stable point $p_L$ in the following two sub-intervals:
1. If $p_S < p_t < p_L$, it can be readily seen from Fig. 4 that $p_{t+1} = \exp(-\hat{\lambda}/p_t) > p_t$. Therefore, $p_t \to p_L$ as $t \to \infty$;
2. If $p_t > p_L$, Fig. 4 shows that $p_{t+1} = \exp(-\hat{\lambda}/p_t) < p_t$. Therefore, $p_t \to p_L$ as $t \to \infty$. □

We can see from Theorem 2 that the unstable equilibrium point $p_S$ is critical for the system stability. As long as the probability of success $p_t$ at time slot $t$ stays above $p_S$, it will ultimately converge to $p_L$, leading to a network throughput of $\hat{\lambda} = -p_L \ln p_L$.

On the other hand, if the probability of success $p_t$ drops below $p_S$, Fig. 4 shows that it will be departing from the desired stable point $p_L$ because $p_{t+1} = \exp(-\hat{\lambda}/p_t) < p_t < p_S$. As the probability of success becomes smaller and smaller, all the nodes will eventually become busy and the network is saturated, in which case the multinomial distribution assumption adopted in the proof of Theorem 1 is no longer valid. As a consequence, the probability of success will not be governed by the characteristic equation (9) displayed in Fig. 4. Instead of zero, $p_t$ will converge to an undesired stable point $p_A$ that depends on the backoff protocols. A detailed discussion on the undesired stable point will be presented in Section V.

As shown in Fig. 5, Theorem 2 can also be paraphrased in terms of attempt rate $G_t$ at time slot $t$: If $G_t$ is to the left of the unstable equilibrium point $-W_{-1}(-\hat{\lambda})$, then $G_t$ converges to the desired stable point $-W_0(-\hat{\lambda})$ as $t \to \infty$.

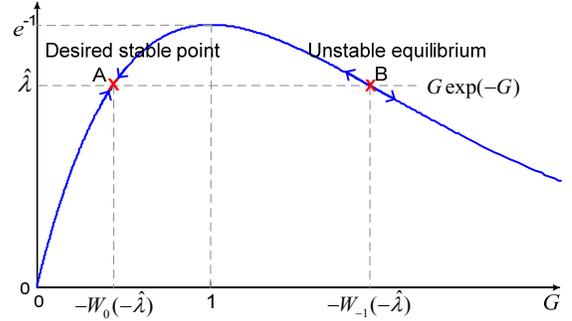

Fig. 5. Stable points of attempt rate $G$

## III. ABSOLUTE STABILITY

The core issue of stability analysis is to characterize the range of retransmission factor $q$, inside which the steady-state network throughput $\hat{\lambda} = n\lambda$ can be achieved. In this section, we will investigate the conditions of *absolute stability* that guarantee the convergence of probability of success to the desired stable point $p_L$ for achieving the throughput $\hat{\lambda}$.

It has been shown in Theorem 2 that the convergence of probability of success to $p_L$ requires that
$$p_t \geq p_S = \exp\{W_{-1}(-\hat{\lambda})\}, \quad (18)$$
or equivalently, the attempt rate $G_t$ satisfies
$$G_t \leq -\ln p_S = -W_{-1}(-\hat{\lambda}) \quad (19)$$
at any time slot $t$. In the next theorem, we will show that this constraint imposes an upper bound on the retransmission factor $q$.

**Theorem 3.** *For K-exponential backoff ($1 \leq K \leq \infty$), if*
$$q \leq q_u = -\ln p_S / n = -W_{-1}(-\hat{\lambda})/n, \quad (20)$$
*then at any time slot t, $G_t \leq -\ln p_S$.*

Proof: Suppose that there are totally $n_b$ backlogged HOL packets at time slot $t$, with $n_i$ packets in phase $i$, $i=1,\ldots,K$. We have $\sum_{i=1}^{K} n_i = n_b$. The attempt rate $G_t$ is then given by
$$G_t = (n-n_b)\lambda + \sum_{i=1}^{K} n_i q^i \leq (n-n_b)\lambda + n_b q \quad (21)$$
where the right side of (21) is the attempt rate corresponding to the state that the $n_b$ backlogged HOL packets are all in phase 1. We consider the following two cases:
1) If retransmission factor $q \leq \lambda$, the attempt rate $G_t$ is bounded by

$$G_t \leq \hat{\lambda}. \tag{22}$$

We know from (17) that

$$-\ln p_S = -W_{-1}(-\hat{\lambda}) > \hat{\lambda}. \tag{23}$$

By combining (22) and (23), we have

$$G_t \leq -\ln p_S. \tag{24}$$

2) If retransmission factor $q \geq \lambda$, the attempt rate $G_t$ is bounded by

$$G_t \leq nq \leq nq_u = -\ln p_S. \tag{25}$$

Hence, the theorem is established by combining (24) and (25). □

Another vital criterion imposed on the range of retransmission factor $q$ is that the offered load $\rho$ of each input queue given by (6) must be less than 1 to ensure that the queue length of each Geo/G/1 queue is bounded. The lower bound of the retransmission factor $q$, specified in the following theorem, is derived from the offered load $\rho$.

**Theorem 4.** *For K-exponential backoff ($1 \leq K \leq \infty$), suppose the probability of success converges to $p_L$, then $\rho \leq 1$ iff $q \geq q_l$, where the lower bound $q_l$ is the root of the following equation:*

$$\frac{q}{p_L + q - 1} - \left( \frac{q}{p_L + q - 1} - \frac{1}{p_L} \right) \cdot \left( \frac{1 - p_L}{q} \right)^K = 1/\lambda. \tag{26}$$

*In particular, the lower bound $q_l$ for geometric retransmission (K=1) is*

$$q_l^{Geo} = \frac{\lambda(1 - p_L)}{p_L(1 - \lambda)} = \frac{\hat{\lambda}(1 - p_L)}{p_L(n - \hat{\lambda})}, \tag{27}$$

*and for exponential backoff ($K=\infty$), we have*

$$q_l^{Exp} = \frac{1 - p_L}{1 - \lambda} = \frac{1 - p_L}{1 - \hat{\lambda}/n}. \tag{28}$$

Proof: As the probability of success $p$ converges to $p_L$, it is easy to show from (6) that the offered load $\rho$ monotonically increases with $(1-p_L)/q$, and therefore $\rho$ is a monotonic decreasing function of $q$. It follows that the minimum retransmission factor $q$ corresponds to the maximum offered load $\rho$. Thus, the lower bound of $q$ is the root of $\rho=1$, which transforms (6) into (26). □

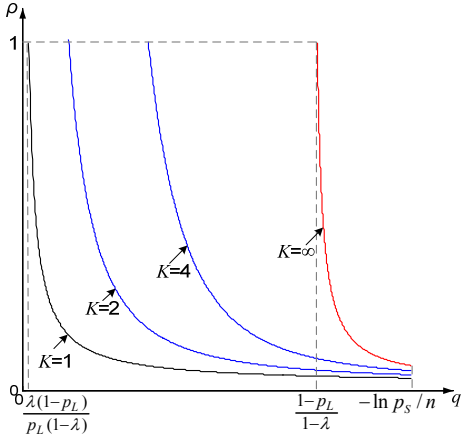

Fig. 6. Tradeoff between offered load $\rho$ and retransmission factor $q$.

Note that the retransmission factor $q$ should be strictly larger than $q_l$ if bounded queue length is required. Besides, according to Theorems 2 and 3, $q$ should not exceed $q_u$ to guarantee that the probability of success converges to $p_L$. The offered load $\rho$ versus the retransmission factor $q$ is plotted in Fig. 6 under different values of cutoff phase $K$. It shows that the offered load $\rho$ is a monotonic decreasing function of $q$ for any given $K$. Fig. 6 also indicates that a larger cutoff phase $K$ leads to a higher offered load $\rho$, which incurs a larger delay, for any given retransmission factor $q$.

Based on Theorems 3 and 4, we define the *absolute stable region* of retransmission factor $q$ as $S = [q_l, q_u]$. According to (20) and (27), the absolute stable region of geometric retransmission is given by

$$S^{Geo} = \left[ q_l^{Geo}, q_u^{Geo} \right] = \left[ \frac{\hat{\lambda}(1-p_L)}{p_L(n-\hat{\lambda})}, -\ln p_S / n \right]. \tag{29}$$

For exponential backoff, from (20) and (28), we have

$$S^{Exp} = \left[ q_l^{Exp}, q_u^{Exp} \right] = \left[ \frac{1-p_L}{1-\hat{\lambda}/n}, -\ln p_S / n \right]. \tag{30}$$

If the retransmission factor is chosen from the absolute stable region, $q \in S = [q_l, q_u]$, the system will stabilize at the desired stable point $p_L$ for sure. The stable region $S$ is determined by the aggregate input rate $\hat{\lambda}$ and the total number of nodes $n$. Apparently, the region $S$ will become an empty set if either the aggregate input rate $\hat{\lambda}$ is too large or the number of nodes $n$ goes to infinity. For any fixed $n$, the maximum stable throughput, denoted by $\hat{\lambda}_{max\_S} = \sup_S \hat{\lambda}$, can be determined as follows.

**Lemma 2.** *For networks with n nodes, the maximum stable throughput $\hat{\lambda}_{max\_S}$ is the root of the following equation:*

$$q_l(\hat{\lambda}) - q_u(\hat{\lambda}) = 0, \tag{31}$$

*with retransmission factor $q = q_l(\hat{\lambda}_{max\_S}) = q_u(\hat{\lambda}_{max\_S})$.*

Proof: Since $q_l$ and $q_u$ are monotonic increasing and decreasing functions, respectively, of the aggregate input rate $\hat{\lambda}$ for any fixed $n$, the maximum stable throughput must be the single root of (31). □

Note that the offered load $\rho$ of each input queue will become 1 if the aggregate input rate is $\hat{\lambda}_{max\_S}$, and the queue length will be unbounded. Thus, the maximum stable throughput $\hat{\lambda}_{max\_S}$ determined by (31) simply means a network throughput of $\hat{\lambda}_{max\_S} - \varepsilon$, for any positive number $\varepsilon > 0$, is achievable with some retransmission factor $q \in S$.

The absolute stable region $S^{Geo}$ of geometric retransmission is depicted in Fig. 7. The region becomes narrower and narrower as the aggregate input rate $\hat{\lambda}$ increases. It eventually shrinks to a single point at $\hat{\lambda} = \hat{\lambda}_{max\_S}^{Geo}$, which is given in the next corollary.

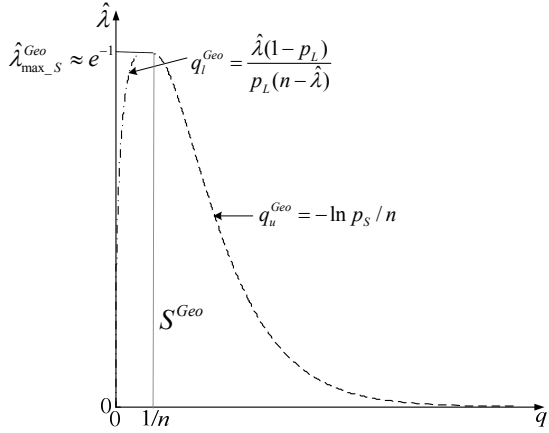

Fig. 7. Absolute stable region and maximum stable throughput of geometric retransmission ($K=1$).

**Corollary 1**. *For geometric retransmission ($K=1$), the maximum stable throughput is approximately given by*

$$\hat{\lambda}_{max\_S}^{Geo} \approx e^{-1} \qquad (32)$$

*with retransmission factor $q=1/n$.*

Proof: According to (29), the difference between the upper bound $q_u^{Geo}$ and the lower bound $q_l^{Geo}$ can be approximately given by $q_u^{Geo}(\hat{\lambda}) - q_l^{Geo}(\hat{\lambda}) \approx \frac{1}{n}\left(-\ln p_S - \hat{\lambda}(1-p_L)/p_L\right)$, which is a monotonic decreasing function of $\hat{\lambda}$. When $\hat{\lambda}=e^{-1}$, the difference is $e^{-1}/n$, which approaches zero when the number of nodes $n$ is large. Thus, the assertion of this corollary directly follows from Lemma 2. □

Both the upper bound $q_u^{Geo}$ and lower bound $q_l^{Geo}$ will approach zero, and the absolute stable region of geometric retransmission will vanish, $S^{Geo}=\varnothing$, when the number of nodes $n\to\infty$. As the retransmission factor $q=0$ is the only viable choice, a node will be unable to transmit any packets once it is involved in a collision. It reveals the fact that the network cannot be stabilized at the point $p_L$ and the throughput is undetermined when the number of nodes $n$ is infinite.

The absolute stable region $S^{Exp}$ depicted in Fig. 8 shows that the maximum stable throughput of exponential backoff could be lower than $e^{-1}$. We prove in the next corollary that it is the case for any given number of nodes $n$.

**Corollary 2**. *For exponential backoff ($K=\infty$), the maximum stable throughput is approximately given by*

$$\hat{\lambda}_{max\_S}^{Exp} \approx \ln n/n \qquad (33)$$

*with retransmission factor $q\approx\ln n/n$.*

Proof: It is known from (15) that $1-\hat{\lambda}$ is a good approximation for $p_L$. Therefore, the lower bound $q_l^{Exp}$ can be approximated by

$$q_l^{Exp} = \frac{1-p_L}{1-\hat{\lambda}/n} \approx \frac{\hat{\lambda}}{1-\hat{\lambda}/n} \approx \hat{\lambda} \qquad (34)$$

when the number of nodes $n$ is large. According to Lemma 2, the maximum throughput $\hat{\lambda}_{max\_S}^{Exp}$ satisfies the following equation obtained by combining (20) and (34):

$$\hat{\lambda} = \frac{-W_{-1}(-\hat{\lambda})}{n}. \qquad (35)$$

Solving (35), we can obtain (33) and the corresponding retransmission factor $q\approx \hat{\lambda}_{max\_S}^{Exp} \approx \ln n/n$. □

Note that the exact value of $\hat{\lambda}_{max\_S}^{Exp}$ should be lower than $\ln n/n$ as $q_l^{Exp}$ is strictly larger than $\hat{\lambda}$. Nevertheless, this approximation is quite accurate for large $n$.

In contrast to geometric retransmission, here the maximum stable throughput $\hat{\lambda}_{max\_S}^{Exp} \approx \ln n/n$ decreases rapidly with the increasing number of nodes $n$. Again, the stable region $S^{Exp}$ becomes an empty set when the number of nodes $n\to\infty$.

In a finite-node network, however, the system can be stabilized at $p_L$ if the aggregate input rate is lower than $\hat{\lambda}_{max\_S}^{Exp} \approx \ln n/n$. This result agrees with that reported in [14-16], which shows that the network with binary exponential backoff (BEB) is stable if the arrival rate is sufficiently small. This point is clearly illustrated in Fig. 8. It is easy to show that if the aggregate input rate $\hat{\lambda}$ is lower than $\frac{1}{2}ne^{-n/2}$, then the retransmission factor $q=1/2$ is included in the stable region $S^{Exp}$.

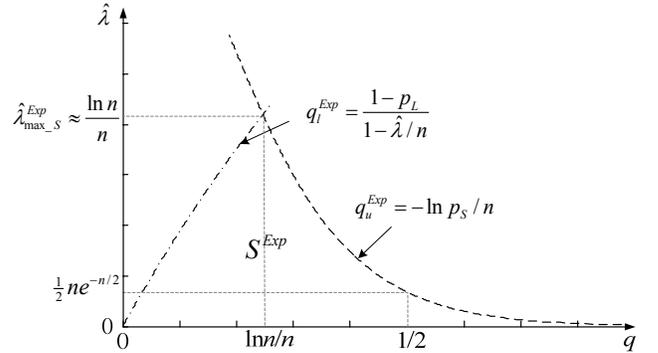

Fig. 8. Absolute stable region and maximum stable throughput of exponential backoff ($K=\infty$).

For the general $K$-exponential backoff with $1<K<\infty$, it is quite difficult to obtain the explicit expression of the lower bound $q_l$. In the Appendix we show that with a large number of nodes $n$, the lower bound $q_l$ is approximately given by

$$q_l = \frac{1-p_L}{\sqrt[K]{np_L/\hat{\lambda}}}, \qquad (36)$$

and the maximum stable throughput is

$$\hat{\lambda}_{max\_S} \approx \ln n^{1-1/K}/n^{1-1/K}. \qquad (37)$$

The maximum stable throughput $\hat{\lambda}_{max\_S}$ declines with the increasing cutoff phase $K$, and (37) agrees with (33) when $K=\infty$. We can see from (20) and (36) that both $q_l$ and $q_u$ approach zero as the number of nodes $n$ increases, indicating that the $K$-exponential backoff with $1<K<\infty$ cannot be stabilized at the desired stable point $p_L$ either when the number of nodes is infinite.

The sum and substance of absolute stability is that a network with $K$-exponential backoff can be absolutely stabilized at $p_L$

only if the number of nodes *n* is finite. The maximum stable throughput could be diminished by increasing the cutoff phase *K*. It appears that geometric retransmission is a favorable option for small-scale networks because it yields the highest stable throughput $e^{-1}$. In the next section, however, we will show that absolute stability is overkill for large cutoff phase *K*, and the superiority of exponential backoff is the existence of a larger stable region of retransmission factor *q*, called the *asymptotic stable region*.

## IV. ASYMPTOTIC STABILITY OF EXPONENTIAL BACKOFF

The absolute stable region of retransmission factor described in Section III guarantees that the attempt rate $G_t$ stays below $-\ln p_S$ in the worst state. For exponential backoff, this constraint imposed on the retransmission factor *q* is overly conservative, because there are an infinite number of phases and the occurrence of the worst state that *n* backlogged HOL packets are all in phase 1 is an extremely rare event for large *n*.

In respect to the versatility of exponential backoff with multiple retransmission phases, we consider a better-than-worst scenario that all *n* nodes are backlogged with HOL packets distributed over *K* phases. Let $n_i$, $i=1,\ldots,K$, be the number of HOL packets in phase *i*, such that $\sum_{i=1}^{K} n_i = n$. Let $\phi_i$ denote the probability that an HOL packet is in phase *i* given that it is backlogged. From (3-5) we have

$$\phi_i = f_i/(1-f_0), \ i=1,2,\ldots,K. \quad (38)$$

According to the law of large numbers, the number of HOL packets in phase *i*, $n_i$, will converge to $n_i^* = n\phi_i$, $i=1,\ldots,K$, when the number of nodes *n* is large. The attempt rate, denoted by $G^*$, under the node distribution $\{n_1^*,\ldots,n_K^*\}$ can then be written as

$$G^* = \sum_{i=1}^{K} n_i^* q^i = n\sum_{i=1}^{K} \phi_i q^i . \quad (39)$$

An upper bound $q_u^*$ of retransmission factor that guarantees $G^* \leq -\ln p_S$ should satisfy the following equation:

$$\sum_{i=1}^{K} n_i^* q^i = n\sum_{i=1}^{K} \phi_i q^i = -\ln p_S . \quad (40)$$

The worst case that we have considered in Section III is that all backlogged nodes are in phase 1, corresponding to the node distribution $\{n, 0, \ldots, 0\}$. Thus, we can see from (40) that $q_u^*$ is no less than $q_u$, and it provides a much looser upper bound. For geometric retransmission (*K*=1), the upper bound $q_u^*$ is given by

$$q_u^{*Geo} = -\ln p_S / n = q_u , \quad (41)$$

and for exponential backoff (*K*=∞), we have

$$q_u^{*Exp} = 1 - p_L - \frac{\ln p_S}{n} p_L > q_u . \quad (42)$$

For *K*-exponential backoff with 1<*K*<∞, the upper bound $q_u^*$ is the root of the following polynomial equation with degree *K*+1:

$$\frac{q}{p_L + q - 1} - \left(\frac{q}{p_L + q - 1} - \frac{1}{p_L}\right)\cdot\left(\frac{1-p_L}{q}\right)^K = 1 + \frac{1-p_L}{p_L}\cdot\frac{n}{-\ln p_S}. \quad (43)$$

The following approximation of $q_u^*$ can be obtained by using a similar approach given in the Appendix:

$$q_u^* \approx \frac{1-p_L}{\sqrt[K]{n/\frac{-\ln p_S}{1-p_L}}} \quad \text{for large } n. \quad (44)$$

Certainly, the upper bound $q_u^*$ determined by (40) does not guarantee that the system always stabilizes at the desired stable point $p_L = \exp\{W_0(-\hat{\lambda})\}$. The following theorem shows that the probability that the attempt rate $G_t$ exceeds $-\ln p_S$ is insignificant, and it converges to 0 as *n*→∞.

**Theorem 5.** *For exponential backoff* (*K*=∞), *if the retransmission factor q satisfies*

$$q \in \left[q_l^{Exp}, \ q_u^{*Exp}\right] = \left[\frac{1-p_L}{1-\hat{\lambda}/n}, \ 1-p_L - \frac{\ln p_S}{n} p_L\right], \quad (45)$$

*then for any time slot t, there exists an ε>0 such that*

$$\Pr\{G_t \leq -\ln p_S\} \geq 1-\varepsilon , \quad (46)$$

*and ε→0 as n→∞.*

Proof: Suppose that at time slot *t* there are totally $n_b$ backlogged HOL packets in the network, with $n_i$ packets in phase *i*, *i*=1,2,…. The attempt rate $G_t$ is then given by

$$G_t = (n-n_b)\lambda + \sum_{i=1}^{\infty} n_i q^i = (1-n_b/n)\hat{\lambda} + \sum_{i=1}^{\infty} n_i q^i , \quad (47)$$

where $n_b = \sum_{i=1}^{\infty} n_i$. We then have

$$\Pr\{G_t \leq -\ln p_S\} = \Pr\{\sum_{i=1}^{\infty} n_i q^i \leq -\ln p_S - (1-n_b/n)\hat{\lambda}\}$$
$$\geq 1-\varepsilon , \quad (48)$$

where

$$\varepsilon = \Pr\{\sum_{i=1}^{\infty} n_i q^i > -\ln p_S - (1-n_b/n)\hat{\lambda}\}. \quad (49)$$

The total number of backlogged nodes $n_b$ may vary from 0 to *n*. To complete the proof, we consider two cases, small $n_b$ and large $n_b$, in the following.

1) For small $n_b$, or specifically, $n_b = o(n)$, according to Markov inequality we have

$$\varepsilon \leq \frac{\sum_{i=1}^{\infty} E[n_i] q^i}{-\ln p_S - (1-n_b/n)\hat{\lambda}}. \quad (50)$$

Let $Y_{i,j}$ be a Bernoulli random variable defined by:

$$Y_{i,j} = \begin{cases} 1 & \text{if backlogged node } j \text{ is in phase } i \\ 0 & \text{otherwise} \end{cases}$$

with the probability mass function $\Pr\{Y_{i,j}=1\}=\phi_i$ and $\Pr\{Y_{i,j}=0\}=1-\phi_i$, for each backlogged node *j*. The number of backlogged nodes in phase *i*, $n_i$, *i*=1,2,…, can then be written as $n_i = \sum_{j=1}^{n_b} Y_{i,j}$. Therefore, the mean number of backlogged nodes in phase *i* is $E[n_i] = n_b\phi_i$. According to (50), we obtain

$$\varepsilon \leq \frac{n_b\sum_{i=1}^{\infty} \phi_i q^i}{-\ln p_S - (1-n_b/n)\hat{\lambda}}. \quad (51)$$

The given condition $q \in [q_l^{Exp}, q_u^{*Exp}]$ is equivalent to

$$\frac{(1-p_L)\hat{\lambda}}{p_L(n-\hat{\lambda})} \leq \sum_{i=1}^{\infty} \phi_i q^i \leq \frac{-\ln p_S}{n} , \quad (52)$$

which implies $\sum_{i=1}^{\infty}\phi_i q^i = \Theta(1/n)$. Furthermore, we know that $\lim_{n\to\infty} n_b/n = 0$ for given $n_b = o(n)$. Therefore, from (50) we have

$$\lim_{n\to\infty}\varepsilon \leq \lim_{n\to\infty}\frac{n_b \sum_{i=1}^{\infty}\phi_i q^i}{-\ln p_S - (1-n_b/n)\hat{\lambda}} = 0. \quad (53)$$

2) For large $n_b$, the state $(n_1, n_2, \ldots)$ follows a multinomial distribution with parameters $n_b$ and $(\phi_1, \phi_2, \ldots)$, and the covariance of $(n_i/n_b, n_j/n_b)$, $i \neq j$, given by

$$Cov(n_i/n_b, n_j/n_b) = -\phi_i\phi_j/n_b, \, i \neq j \quad (54)$$

becomes negligible. According to the central limit theorem, the independent random variables $n_1/n_b, n_2/n_b, \ldots$ are normally distributed. The linear combination of $n_i/n_b$, $Z = \frac{1}{n_b}\sum_{i=1}^{\infty} n_i q^i$, also has a normal distribution with mean $\mu_Z = \sum_{i=1}^{\infty}\phi_i q^i$ and variance $\sigma_Z^2 = \frac{1}{n_b}\sum_{i=1}^{\infty}\phi_i(1-\phi_i)q^{2i}$. We have

$$\varepsilon = 1 - \Phi\left(\frac{(-\ln p_S/n_b - (1/n_b - 1/n)\hat{\lambda} + \delta) - \mu_Z}{\sigma_Z}\right), \quad (55)$$

where $\delta > 0$ is a small positive number. We know from (52) that $\mu_Z \leq (-\ln p_S)/n$. Besides,

$$\sigma_Z^2 = \frac{1}{n_b}\sum_{i=1}^{\infty}\phi_i(1-\phi_i)q^{2i} \leq \frac{1}{n_b}\sum_{i=1}^{\infty}\phi_i q^{2i} \leq \frac{1}{n_b}\sum_{i=1}^{\infty}\phi_i q^i. \quad (56)$$

Thus, the probability given in (55) is bounded by

$$\varepsilon \leq 1 - \Phi\left(\frac{\sqrt{n_b}(1/n_b - 1/n)(-\ln p_S - \hat{\lambda}) + \delta}{\sqrt{\sum_{i=1}^{\infty}\phi_i q^i}}\right), \quad (57)$$

where the right side of (57) is maximized when $n_b = n$. Therefore,

$$\varepsilon \leq 1 - \Phi\left(\frac{\delta}{\sqrt{\sum_{i=1}^{\infty}\phi_i q^i/n}}\right). \quad (58)$$

We have shown in (52) that $\sum_{i=1}^{\infty}\phi_i q^i = \Theta(1/n)$. Hence, according to (58), we have

$$\lim_{n\to\infty}\varepsilon \leq 1 - \lim_{n\to\infty}\Phi\left(\frac{\delta}{\sqrt{\sum_{i=1}^{\infty}\phi_i q^i/n}}\right) = 0. \quad (59)$$

From (53) and (59), we conclude that for any $0 \leq n_b \leq n$, $\varepsilon \to 0$ as $n\to\infty$. □

Define $S^* = [q_l^{Exp}, q_u^{*Exp}]$ as the asymptotic stable region of exponential backoff. The maximum asymptotic stable throughput, denoted by $\hat{\lambda}_{max\_S}^* = \sup_{S^*}\hat{\lambda}$, is given in the next corollary.

**Corollary 3**. *For exponential backoff ($K=\infty$), the maximum asymptotic stable throughput is approximately given by*

$$\hat{\lambda}_{max\_S}^* \approx e^{-1} \quad (60)$$

*with retransmission factor $q = 1-e^{-1}$.*

Proof: The proof is similar to that of Corollary 1. According to (45), we have $q_u^{*Exp}(\hat{\lambda}) - q_l^{Exp}(\hat{\lambda}) \approx \frac{-\ln p_S}{n}\cdot p_L$, which is also monotonic decreasing with respect to $\hat{\lambda}$. When $\hat{\lambda} = e^{-1}$, the difference $e^{-1}/n$ approaches zero when the number of nodes $n$ is large. □

The asymptotic stable region $S^*$ is depicted in Fig. 9 under different values of aggregate input rate $\hat{\lambda}$. For the sake of comparison, the absolute stable region $S^{Exp}$ is also plotted. It can be clearly seen that $S^{Exp}$ is a subset of $S^*$ for any $\hat{\lambda}$. Both regions are diminishing with an increase of $\hat{\lambda}$. For $\hat{\lambda} > \hat{\lambda}_{max\_S}^{Exp} \approx \ln n/n$, the region $S^{Exp}$ becomes empty, indicating that the network cannot be absolutely stabilized at the desired stable point $p_L$. Nevertheless, Theorem 5 ensures that it can still be stabilized at $p_L$ with a very high probability, if the retransmission factor $q \in S^*$ and the number of nodes $n$ is large.

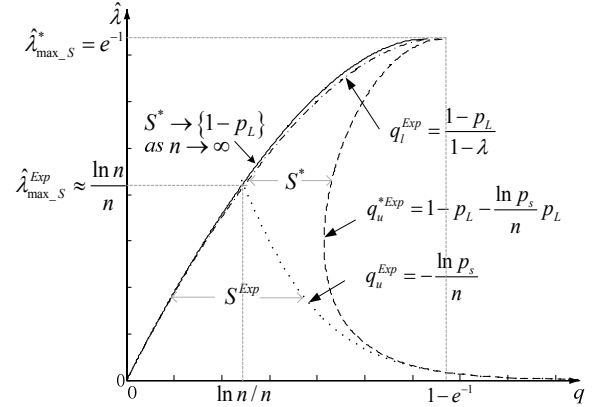

Fig. 9. Asymptotic stable region $S^*$ as a function of aggregate input rate $\hat{\lambda}$ and retransmission factor $q$.

When the number of nodes $n\to\infty$, the absolute stable region $S^{Exp}$ becomes an empty set, and the asymptotic stable region $S^*$ shrinks to a single point $1-p_L$. The corresponding network throughput at this point can be written as

$$\hat{\lambda} = -p_L \ln p_L = -(1-q)\ln(1-q), \quad (61)$$

where $q = 1-p_L$. The maximum throughput $e^{-1}$ is achieved when the retransmission factor $q = 1-e^{-1}$. This is consistent with that claimed in [17], which shows that the network throughput of exponential backoff with an infinite number of nodes is given by $\frac{r-1}{r}\ln\frac{r}{r-1}$, where $r$ corresponds to $1/q$ in our case.

We note that for any given aggregate input rate $\hat{\lambda} \leq e^{-1}$, the input rate of each node $\lambda = \hat{\lambda}/n \to 0$ as $n\to\infty$. Thus, the above result is a natural consequence of Lemma 1 that the probability of success $p_L \to 1-q$ as $n\to\infty$ is the singularity of the mean service time

$$1/f_0 = \rho^{Exp}/\lambda = q/(p_L + q - 1) \to \infty \quad (62)$$

when the offered load $\rho^{Exp} \to 1$. Here, the infinite mean service time simply means that the availability of the server for each individual queue is uncertain in the face of an infinite number of busy nodes in the network. Even though a network throughput of $\hat{\lambda}$ can still be achieved, the delay performance of each input queue would be severely penalized. This point will be further elaborated in the next section on the discussion of pseudo-stability.

With a finite cutoff phase, $K<\infty$, the upper bound $q_u^*$ will approach zero when the number of nodes $n\to\infty$, indicating that no retransmission factor $q$ can be found to make the $K$-exponential backoff with finite $K$ asymptotically stable. In fact, from (44) we can see that the cutoff phase $K$ is used to counteract the effect of the number of nodes $n$ in the $K$-exponential backoff scheme. With an infinite number of nodes, an infinite number of phases is required to stabilize the system at $p_L$.

## V. PSEUDO-STABILITY OF EXPONENTIAL BACKOFF

The preceding analysis focuses on the delineation of stable region of the retransmission factor $q$, in which the system will stabilize at the desired stable point $p_L$. This section is devoted to the analysis of system behavior once the probability of success $p_t < p_S$, and the system may run into the risk of being evolved into the *undesired* stable point.

It is proved in Theorem 2 that the probability of success $p_t$ will monotonically decrease if $p_t < p_S$. Consequently, according to (3), the service rate $f_0$ of each queue will become smaller and smaller and eventually drop below the input rate $\lambda$. In this case, the system becomes saturated and all nodes in the system will be busy with probability 1.

In time slot $t$ when all nodes are busy, for networks with geometric retransmission, all HOL packets will be in phase 1, and the attempt rate is given by $G_t = nq$. For exponential backoff, however, there is no limit on the phases of HOL packets. The nodes can always back off to deeper phases to alleviate contentions, and to make the attempt rate $G_t$ arbitrarily small until the network is stabilized. Thus, exponential backoff is much more adaptive to the traffic fluctuation when compared with geometric retransmission.

To manifest the adaptability of various backoff protocols, we will demonstrate in this section that there exists a non-empty pseudo-stable region for exponential backoff, while geometric retransmission and $K$-exponential backoff with $1<K<\infty$ are inherently unstable if the number of nodes is infinite. We need the following properties of the Lambert W function to facilitate our analysis:

1. Monotonic increasing property of $W_0(z)$:
$$-1 \leq W_0(a) \leq W_0(b) \quad \text{iff} \quad -e^{-1} \leq a \leq b. \quad (63)$$

2. Monotonic decreasing property of $W_{-1}(z)$:
$$-1 \geq W_{-1}(a) \geq W_{-1}(b) \quad \text{iff} \quad -e^{-1} \leq a \leq b \leq 0. \quad (64)$$

The main result of this section is outlined in the following theorem.

**Theorem 6.** *If $p_t < p_S$ at some time slot $t$, the probability of success $p_t$ will converge to $p_A$ as $t \to \infty$, where $p_A$ is the root of the following equation:*

$$p = \exp\left\{-n / \left(\frac{pq}{p+q-1} - \left(\frac{pq}{p+q-1} - 1\right) \cdot \left(\frac{1-p}{q}\right)^K\right)\right\}. \quad (65)$$

*In particular, two special cases are of interest:*
*1) For geometric retransmission ($K=1$), we have*
$$p_A^{Geo} \approx \exp\{-nq\}, \quad (66)$$

*and $p_A^{Geo} \to 0$ as $n \to \infty$.*

*2) For exponential backoff ($K=\infty$), we have*
$$p_A^{Exp} \approx \frac{n(1-q)}{n+q\ln(1-q)}, \quad (67)$$

*and $p_A^{Exp} \to 1-q$ as $n \to \infty$.*

Proof: Similar to the proof of Theorem 1, if all $n$-1 interfering nodes are busy, the probability of success at time slot $t$+1 can be written as

$$p_{t+1} = \left(\sum_{i=1}^{K} f_{i,t}(1-q^i)\right)^{n-1} \overset{\text{with a large } n}{\approx} \exp(-nf_{0,t}/p_t). \quad (68)$$

Substituting (3) into (68), we have
$$p_{t+1} = \exp\{-n/g(p_t)\}, \quad (69)$$

where
$$g(p_t) = \frac{p_t q}{p_t + q - 1} - \left(\frac{p_t q}{p_t + q - 1} - 1\right) \cdot \left(\frac{1-p_t}{q}\right)^K \quad (70)$$

is a monotonic decreasing function of $p_t$. It is straightforward to show that:
i) When $p_{t-1} < p_t$, according to $g(p_{t-1}) > g(p_t)$ we have $p_t > p_{t+1}$;
ii) When $p_{t-1} > p_t$, according to $g(p_{t-1}) < g(p_t)$ we have $p_t < p_{t+1}$.

Therefore, $p_t$ converges to the unique fixed point $p_A$ of (69) as $t \to \infty$. Two specific examples are illustrated as follows.

1) When $K$=1, we have
$$p_A = \exp(-nq/(1-p_A+p_A q)). \quad (71)$$
Since $(1-q)p_A \ll 1$ for a large $n$, we have $p_A \approx \exp\{-nq\} \to 0$ as $n \to \infty$.

2) When $K=\infty$, we have
$$p_A = \exp\left(-n \cdot \frac{q-(1-p_A)}{p_A q}\right). \quad (72)$$

Let $W = n(1/q-1)/p_A$. Then (72) can be written as
$$We^W = n(1/q-1)\exp(n/q). \quad (73)$$

Since $n(1/q-1)\exp(n/q) > 0$, the Lambert W function defined by (73) can be uniquely represented as:
$$W = n(1/q-1)/p_A = W_0(n(1-q)/q \cdot \exp(n/q)). \quad (74)$$

Next, apply (63) and the property $W_0(We^W)=W$ to the following inequality:
$$(n/q-n)\exp(n/q-n) \leq (n/q-n)\exp(n/q) \leq (n/q)\exp(n/q), \quad (75)$$
we immediately obtain
$$n/q-n \leq W_0((n/q-n)\exp(n/q)) \leq n/q. \quad (76)$$

This inequality allows us to assume that
$$W_0((n/q-n)\exp(n/q)) = n/q-x, \quad (77)$$

for some $0 \leq x \leq n$, which is the Lambert W function that satisfies:
$$(n/q-x) \cdot \exp(n/q-x) = (n/q-n)\exp(n/q). \quad (78)$$
Hence, we have
$$\exp(x) = (n/q-x)/(n/q-n). \quad (79)$$

Since $n/q \gg x$ for large $n$, from (79) we have
$$x \approx -\ln(1-q). \quad (80)$$

Finally, we can obtain (67) from (74) by substituting (80) into (77). □

In contrast to the desired stable point $p_L$, which is invariant with respect to the retransmission factor $q$, the undesired stable point $p_A$ is dependent on $q$ and the cutoff phase $K$.

Fig. 10 shows the probability of success $p$ versus retransmission factor $q$ of geometric retransmission. It has been proven in Section III that the probability of success converges at $p_L$ with a retransmission factor $q \in S^{Geo}$. However, it drops significantly below $p_L$ when $q > q_u^{Geo} = -\ln p_S/n$. As the number of nodes $n \to \infty$, both $q_l^{Geo}$ and $q_u^{Geo}$ approach zero and the stable region $S^{Geo}$ becomes empty. Besides, the undesired stable point $p_A^{Geo} \to 0$ for any given retransmission factor $q$. Therefore, we can conclude that the geometric retransmission system is unstable if the number of nodes is infinite. This result is consistent with the previous studies [4-5] that the slotted Aloha network with geometric retransmission is inherently *unstable* as $n \to \infty$.

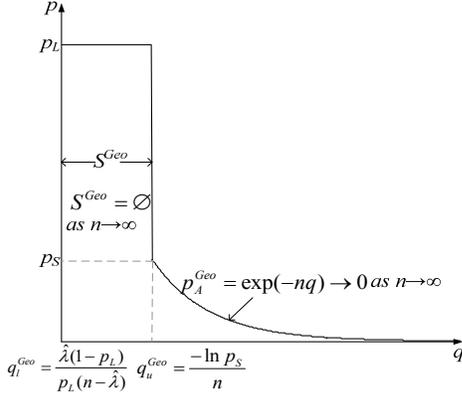

Fig. 10. Probability of success $p$ versus retransmission factor $q$ in the geometric retransmission case ($K$=1).

For exponential backoff, the curve of probability of success $p$ versus retransmission factor $q$ is shown in Fig. 11. It has been proven in Section IV that the probability of success asymptotically converges to $p_L$ if $q \in S^*$. We note that the probability of success almost linearly decreases with respect to $q$ outside the asymptotic stable region $S^*$. Furthermore, because the asymptotic stable region $S^* \to \{1-p_L\} \neq \emptyset$ and the probability of success $p_A^{Exp} \to 1-q$, as $n \to \infty$, the exponential backoff system may remain stable even when the retransmission factor $q$ exceeds the region $S^*$. This point will be further elaborated via the throughput analysis.

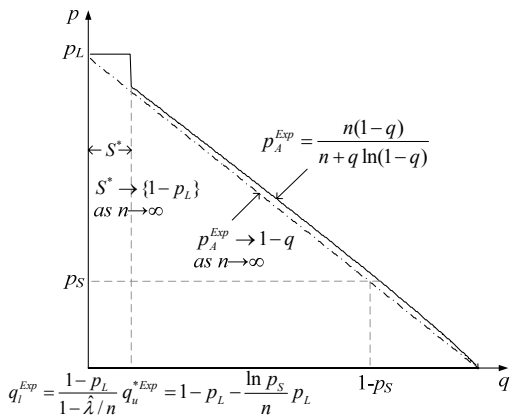

Fig. 11. Probability of success $p$ versus retransmission factor $q$ in the exponential backoff case ($K=\infty$)

The stability analysis in respect to the throughput is based on the comparison of service rate $f_0$ and input rate $\lambda$. Specifically, the network throughput is given by:

$$\hat{\lambda}_{out} = \min\{nf_0, \hat{\lambda}\}. \quad (81)$$

The network will be unstable if $\hat{\lambda}_{out} = nf_0 < \hat{\lambda}$, in which case some input packets will be dropped with a finite buffer at each node, or the queue length may become unbounded if the buffer is infinite.

**Corollary 4**. *For geometric retransmission (K=1), the network throughput $\hat{\lambda}_{out} = \hat{\lambda}$ iff $q \in S^{Geo}$.*

Proof: If $q \in S^{Geo}$, we know from Section III that the network is stable and $\hat{\lambda}_{out} = \hat{\lambda}$. Otherwise, we consider two cases:

1) When the retransmission factor $q > q_u^{Geo} = -\ln p_S/n$, the probability of success will drop below $p_S$ and converge to $p_A^{Geo}$ according to Theorem 6. The corresponding service rate $f_0$ of each single queue satisfies:

$$nf_0 = \frac{np_A^{Geo}q}{1-p_A^{Geo}+p_A^{Geo}q} \approx nq\exp(-nq). \quad (82)$$

Since $nq > -\ln p_S \geq 1$, (82) suggests that $nq$ can be represented by the Lambert W function $-W_{-1}(-nf_0)$, and it follows from (64) that the given condition

$$-W_{-1}(-nf_0) = nq > -\ln p_S = -W_{-1}(-\hat{\lambda}) \geq 1 \quad (83)$$

implies $nf_0 < \hat{\lambda}$. According to (81), the network throughput is given by

$$\hat{\lambda}_{out} = nf_0 = nq\exp\{-nq\} < \hat{\lambda}. \quad (84)$$

2) If $q < q_l^{Geo}$, according to Theorem 4 the service rate $f_0$ will be lower than the input rate $\lambda$. Therefore, we have $\hat{\lambda}_{out} = nf_0 < \hat{\lambda}$. □

For exponential backoff, however, Theorem 6 shows that if the retransmission factor is outside the stable region, $q \notin S^*$, the undesired stable point $p_A^{Exp}$ does not converge to zero even in a network with an infinite number of nodes. The next corollary manifests that exponential backoff is much more robust than geometric retransmission when the network is saturated, and there is a stable region larger than $S^*$ in which a network throughput of $\hat{\lambda} \leq e^{-1}$ is achievable.

**Corollary 5**. *For exponential backoff ($K=\infty$), if the probability of success converges to $p_A^{Exp}$, we have*

$$\hat{\lambda}_{out} = \begin{cases} \hat{\lambda} & \text{if } 1-p_L \leq q \leq 1-p_S \\ -(1-q)\ln(1-q) < \hat{\lambda} & \text{otherwise} \end{cases}$$

Proof: If the probability of success converges to $p_A^{Exp}$, it follows from (67) that the service rate $f_0$ of each single queue satisfies:

$$f_0 = \frac{p_A^{Exp}+q-1}{q} \approx \frac{-(1-q)\ln(1-q)}{n+q\ln(1-q)} \approx -(1-q)\ln(1-q)/n. \quad (85)$$

The two roots of (85) are then given by Lambert W function as follows:

$$1-q = \begin{cases} e^{W_0(-nf_0)} \geq e^{-1} \\ e^{W_{-1}(-nf_0)} \leq e^{-1} \end{cases} \quad (86)$$

The given condition $1-p_L \leq q \leq 1-p_S$ is equivalent to
$$e^{W_{-1}(-\hat{\lambda})} = p_S \leq 1-q \leq p_L = e^{W_0(-\hat{\lambda})}. \quad (87)$$
The combination of (86) and (87) yields
$$e^{W_{-1}(-\hat{\lambda})} \leq e^{W_{-1}(-nf_0)} \leq e^{-1} \leq e^{W_0(-nf_0)} \leq e^{W_0(-\hat{\lambda})}. \quad (88)$$
It follows from (63-64) that (88) implies $nf_0 \geq \hat{\lambda}$. According to (81), the network throughput is $\hat{\lambda}$. On the other hand, if the retransmission factor satisfies either $q<1-p_L$ or $1-p_S<q$, using the similar procedure we can prove that $nf_0 < \hat{\lambda}$. In this case, the network throughput is $\hat{\lambda}_{out} = nf_0 = -(1-q)\ln(1-q) < \hat{\lambda}$. □

Despite its robustness, a network with exponential backoff may suffer from severe delay jitter when the probability of success drops below $p_S$ and converges to $p_A^{Exp}$. The nodes would have to back off to much deeper phases with extremely small retransmission probabilities when saturation occurs. As a result, once a node tries to retransmit and succeeds, it is very likely that this node will dominate the channel for a fairly long period of time and produce a continuous stream of packets until it is interrupted by the retransmission requests initiated by other backlogged nodes. This "capture phenomenon" occurred when the network becomes saturated has been described in [30].

The reason that the channel could be captured by a single node during saturation is twofold: After the first packet being successfully transmitted, subsequent packets in the queue are all in phase 0 and can be transmitted immediately once they are being moved to the HOL position. At the same time, the contentions from all other nodes are insignificant due to their low retransmission probabilities.

The exponential backoff protocol actually serves as an adaptive traffic regulator of the shared channel. When the network is saturated, a stable throughput is achieved by clearing the backlogged packets in one node's queue with very little interferences from the other nodes. The channel takes random turns to serve these backlogged nodes, whose output processes may not be stationary any longer because of the capture effect.

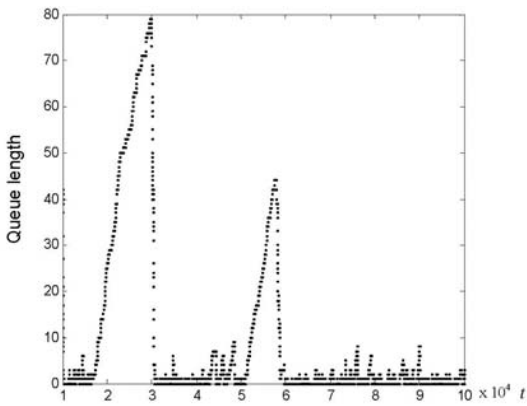

Fig. 12. Non-stationary queue length of a single node in a network with exponential backoff ($K=\infty$), where $n$=50, $\hat{\lambda}$=0.3 and $q$=0.8.

We know from Theorem 6 that in a saturated network with an infinite number of nodes $n$, the undesired stable point $p_A^{Exp} \rightarrow 1-q$ is the singularity of the offered load $\rho^{Exp}=\lambda q/(p+q-1)$. Thus, the range $\{q \mid 1-p_L \leq q \leq 1-p_S\}$ is called the *pseudo-stable region*. In this region, as the probability of success drops below $p_S$ and converges to the undesired stable point $p_A^{Exp}$, the network throughput $\hat{\lambda}$ can still be achieved at the cost of severe delay jitter, resulting unpredictable queuing behavior. The simulation result shown in Fig. 12 demonstrates the non-stationary queue length caused by the capture effect of a single node, which may not have a steady-state distribution.

For $K$-exponential backoff with $1<K<\infty$, the network can yield higher throughput with a larger cutoff phase $K$ outside the stable region. Similar to the case of asymptotic stability, a non-empty pseudo-stable region cannot be warranted by the $K$-exponential backoff with $1<K<\infty$ as $n\rightarrow\infty$. Since the explicit expression of $p_A$ for a general $K$ is rather complicated, we can only sketch some distinguishing characteristics of $p_A$ in the following corollary.

**Corollary 6.** *For K-exponential backoff with $1<K<\infty$, if $p_t < p_S$ at some time slot t, as $t\rightarrow\infty$, the limiting probability of success $p_A$ has the following properties:*
*1) $p_A$ is monotonic increasing with respect to cutoff phase K.*
*2) For any given retransmission factor q, $p_A\rightarrow 0$ as $n\rightarrow\infty$.*

Proof: 1) It can be seen from (70) that for any given $p_t$ and $q$, $g(p_t)$ is monotonic increasing with respect to the cutoff phase $K$. Suppose $K_1<K_2$. Let $p_{A,1}$ and $p_{A,2}$ represent the corresponding convergent points of $p_t$. If $p_t<p_S$, according to (69) in the proof of Theorem 6, we then have
$$p_{A,1} = \exp\{-n/g_{K_1}(p_{A,1})\}, \quad (89)$$
and
$$p_{A,2} = \exp\{-n/g_{K_2}(p_{A,2})\}, \quad (90)$$
Combining (89) and (90), we have
$$g_{K_2}(p_{A,2})\ln\tfrac{1}{p_{A,2}} = n = g_{K_1}(p_{A,1})\ln\tfrac{1}{p_{A,1}} < g_{K_2}(p_{A,1})\ln\tfrac{1}{p_{A,1}}. \quad (91)$$
Since both $g(p)$ and $\ln(1/p)$ are monotonic decreasing functions of $p$, the inequality (91) can only be held for $p_{A,1} < p_{A,2}$. It follows that $p_A$ is monotonic increasing with respect to $K$.

2) According to (65), we know that $p_A$ satisfies
$$p_A = \exp\left\{-n/\left(\frac{p_A q}{p_A+q-1} - \left(\frac{p_A q}{p_A+q-1}-1\right)\cdot\left(\frac{1-p_A}{q}\right)^K\right)\right\}. \quad (92)$$
Let $x=(1-p_A)/q$, and rewrite (92) as
$$-\ln(1-qx)\cdot\left((1-qx)\frac{1-x^K}{1-x}+x^K\right) = n. \quad (93)$$

For any finite cutoff phase $K$, it is clear from (93) that $x\rightarrow 1/q$ as $n\rightarrow\infty$. Therefore, the probability of success $p_A\rightarrow 0$ as $n\rightarrow\infty$. □

We have shown in Section III that the stable region becomes empty for any finite cutoff phase $K$ as $n\rightarrow\infty$. The above corollary further indicates that even $p_A$ can be improved with a larger cutoff phase $K$, the $K$-exponential backoff with $1<K<\infty$ remains *unstable* if the number of nodes is infinite.

## VI. SIMULATION RESULTS AND DISCUSSIONS

In this section, we will provide the simulation results corresponding to preceding theoretical analyses. We first consider a small network ($n$=10 nodes) with light traffic (the aggregate input rate $\hat{\lambda}$=0.1). Fig. 13 presents the curves of the offered load $\rho$ versus the retransmission factor $q$ under different values of cutoff phase $K$. The simulation results perfectly agree with (6) in the stable region of $q$. For the sake of clarity, in the following figures we only provide the corresponding curves of geometric retransmission ($K$=1) and exponential backoff ($K$=∞).

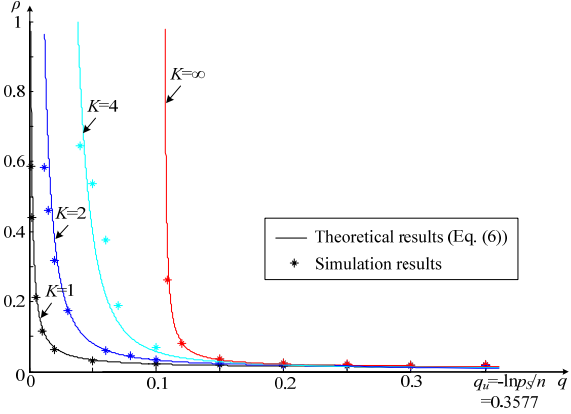

Fig. 13. Offered load $\rho$ versus retransmission factor $q$ with $n$=10 and $\hat{\lambda}$ = 0.1.

The simulation results shown in Fig. 14 clearly demonstrate that both probability of success $p$ and attempt rate $G$ are invariant with respect to retransmission factor $q$ and cutoff phase $K$ in the stable region of $q$. It is a solid confirmation of Theorem 1. We have proved in Theorem 3 that the system should stabilize at the desired stable point $p_L = \exp\{W_0(-\hat{\lambda})\}$ with a retransmission factor $q$ chosen from the stable region $S$. This point is also verified by the simulation results displayed in Fig. 14, which depicts the curves of the probability of success $p$, the attempt rate $G$ and the network throughput in steady state.

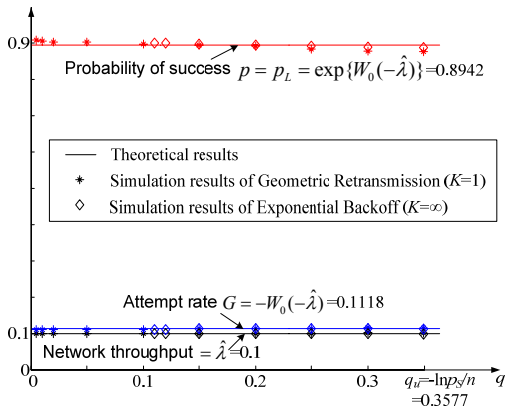

Fig. 14. Probability of success $p$, attempt rate $G$ and network throughput versus retransmission factor $q$ with $n$=10 and $\hat{\lambda}$ = 0.1.

The stable region $S$ will be rapidly diminishing with an increase of the number of nodes $n$ or the aggregate input rate $\hat{\lambda}$. For example, when $n$=50 and $\hat{\lambda}$=0.3, the stable region of geometric retransmission $S^{Geo}$ becomes [0.0038, 0.0356] according to (29). With a retransmission factor $q$ outside the stable region $S^{Geo}$, the probability of success significantly drops from $p_L$ and sharply decays with $q$. This point is verified by the curves depicted in Fig. 15. With exponential backoff, the probability of success decreases, approximately linearly, with the retransmission factor $q$ when $q$ is outside the asymptotic stable region $S^*$=[0.3893, 0.4088], which is given by (45), while it stays at $p_L$ inside the region $S^*$.

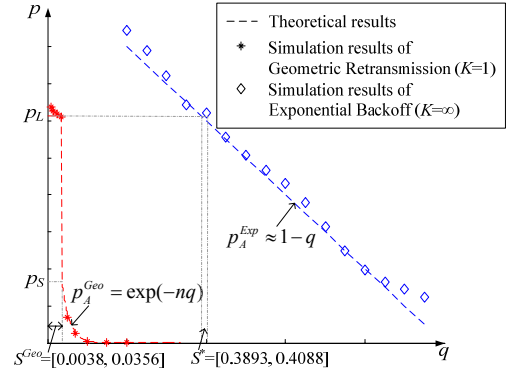

Fig. 15. Probability of success $p$ versus retransmission factor $q$ with $n$=50 and $\hat{\lambda}$ = 0.3.

Fig. 16 shows that the network throughput of geometric retransmission quickly approaches zero when the retransmission factor $q$ exceeds the stable region $S^{Geo}$. This verifies the theoretical analysis in Sections III and V. Our analyses also demonstrate that exponential backoff is rather robust compared with geometric retransmission. It can be readily seen from Fig. 16 that a network throughput of $\hat{\lambda}$=0.3 is achievable if the retransmission factor $q$ is chosen from the range of 0.387=1-$p_L$≤$q$≤1-$p_S$=0.8316. Outside the pseudo-stable region, the network is unstable, and therefore becomes less predictable by analysis. The discrepancy shown in Fig 16 when $q$>1-$p_S$ or $q$<1-$p_L$ mainly stems from the non-stationary queue length of each individual node when the network is saturated.

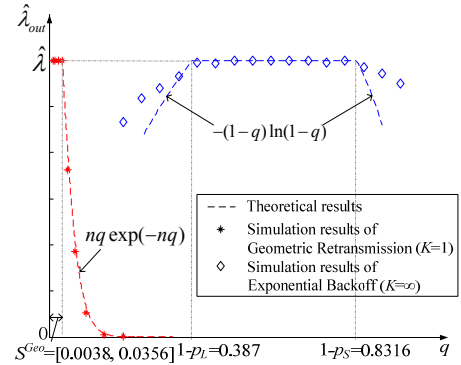

Fig. 16. Network throughput versus retransmission factor $q$ with $n$=50 and $\hat{\lambda}$ = 0.3.

The simulation results presented in this section reinforce the adaptability of exponential backoff that has been investigated in Section V. It outperforms geometric retransmission in terms of network throughput outside the absolute stable region when the number of nodes is large. As we have mentioned before, the main drawback of the pseudo-stability attributes to the potential unbounded delay that could be experienced by packets waiting in the input buffers.

## VII. Conclusions

In this paper, a unified approach is exploited to determine the stable region and throughput of a buffered Aloha network with exponential backoff collision resolution algorithms. For networks with a finite number of nodes, an absolute stable region can be determined for any $K$-exponential backoff that guarantees the network to converge to the desired stable point. Our analyses show that the maximum throughput in this region diminishes with increasing cutoff phase $K$. Thus, geometric retransmission ($K=1$) could be a favorable option for small-scale networks. The requirement of absolute convergence is overkill for large cutoff phase $K$. We show that there exists a bigger asymptotic stable region for exponential backoff ($K=\infty$), which guarantees that the network will be stabilized at the desired stable point with a very high probability. When compared to geometric retransmission, exponential backoff is more resilient in dealing with transient fluctuations of traffic. For networks with an infinite number of nodes, geometric retransmission is unstable, while exponential backoff can be pseudo-stable.

Several interesting topics deserve future research, including the delay analysis of input queue, the compatibility of the cutoff phase $K$ and the number of nodes $n$ to strike a balance between network throughput and queuing delay. Our approach can be extended to explore other collision resolution protocols such as CSMA.

## Appendix. Derivation of $q_l$ and $\hat{\lambda}_{\max\_S}$ of $K$-Exponential Backoff ($1<K<\infty$).

The lower bound $q_l$ is the root of equation $\rho=1$. Let $x=(1-p_L)/q$. According to (6) the offered load $\rho$ can be written as

$$\rho=\lambda\left(\frac{1}{1-x}-\left(\frac{1}{1-x}-\frac{1}{p_L}\right)\cdot x^K\right)=\lambda\left(\frac{1-x^K}{1-x}+\frac{1}{p_L}x^K\right). \quad (94)$$

Suppose that $x^*$ is the solution of equation:

$$p_L\frac{1-x^K}{1-x}+x^K=\frac{n}{\hat{\lambda}}\cdot p_L. \quad (95)$$

With a large $n$, $x^*>1$, we have

$$p_L\frac{1-x^K}{1-x}+x^K\approx x^K. \quad (96)$$

Substituting (96) into (95), we obtain $x^*$ approximately as follows

$$x^*\approx(np_L/\hat{\lambda})^{1/K}. \quad (97)$$

The lower bound $q_l$ is therefore given by

$$q_l=(1-p_L)/x^*\approx(1-p_L)/\sqrt[K]{np_L/\hat{\lambda}}. \quad (98)$$

When $K$ is large, we have

$$q_l\approx\frac{1-p_L}{\sqrt[K]{np_L/\hat{\lambda}}}\approx\frac{\hat{\lambda}^{1+1/K}}{\sqrt[K]{n}}\approx\frac{\hat{\lambda}}{\sqrt[K]{n}}. \quad (99)$$

According to Lemma 2, the maximum stable throughput $\hat{\lambda}_{\max\_S}$ can be obtained by combining (20) and (99):

$$\hat{\lambda}/n^{1/K}=-\ln p_S/n \Rightarrow \hat{\lambda}_{\max\_S}\approx\ln n^{1-1/K}/n^{1-1/K}. \quad (100)$$

## References

[1] N. Abramson, "The Aloha System – Another Alternative for Computer Communication," *Proc. Fall Joint Compet. Conf.*, AFIP Conference, vol. 44, pp. 281-285, 1970.
[2] A. Ephremides and B. Hajek, "Information theory and communication networks: An unconsummated union," *IEEE Trans. Inf. Theory*, vol. 44, no. 6, pp. 2416-2434, Oct. 1998.
[3] L. Tong, V. Naware, and P. Venkitasubramaniam, "Signal Processing in Random Access, " *IEEE Signal Processing Magazine*, Sept 2004.
[4] G. Fayolle, E. Gelenbe, and J. Labetoulle, "Stability and optimal control of the packet switching broadcast channle," *J. Assoc. Comput. Machinery*, vol. 24, pp. 375-386, July 1977.
[5] W. A. Rosenkrantz and D. Towsley, "On the instability of the slotted Aloha multiaccess algorithm," *IEEE Trans. Automat. Contr.*, vol. 28, no. 10, pp. 994-996, Oct. 1983.
[6] A. B. Carleial and M. E. Hellman, "Bistable behavior of Aloha-type systems," *IEEE Trans. Commun.*, vol. 23, no. 4, pp. 401-410, Apr. 1975.
[7] Y. –C. Jenq, "On the stability of slotted Aloha systems," *IEEE Trans. Commun.*, vol. 28, no. 11, pp. 1936-1939, Nov. 1980.
[8] S. Lam and L. Kleinrock, "Packet switching in a multi-access broadcast channel: Dynamic control procedures," *IEEE Trans. Commun.*, vol. COM-23, pp. 891-904, 1975.
[9] B. E. Hajek and T. van Loon, "Decentralized dynamic control of a multiaccess broadcast channel," *IEEE Trans. Automat. Contr.*, vol. 27, pp. 559-569, June 1982.
[10] V. A. Mikhailov, *Methods of Random Multiple Access*, Candidate Engineering Thesis, Moscow Institute of Physics and Technology, Moscow, 1979.
[11] R. Rivest, "Network control by Bayesian broadcast," *IEEE Trans. Inf. Theory*, vol. IT-33, pp. 323-328, May 1987.
[12] R. M. Metcalfe and D. R. Boggs, "Ethernet: distributed packet switching for local computer networks," *Commun. ACM*, pp. 395-404, Jul. 1976.
[13] D. Aldous, "Ultimate instability of exponential back-off protocol for acknowledgement-based transmission control of random access communication channels, " *IEEE Trans. Inf. Theory*, pp 219-223, 1987.
[14] J. Goodman, A. G. Greenberg, N. Madras and P. March, "Stability of binary exponential backoff," *J. ACM*, vol. 35, pp. 579-602, 1988.
[15] J. Hastad, T. Leighton and B. Rogoff, "Analysis of backoff protocols for multiple access channels," *SIAM J. Comput.*, vol. 25, pp. 740-744, 1996.
[16] H. AL-Ammal, L. A. Goldberg and P. MacKenzie, "An improved stability bound for binary exponential backoff," *Theory Comput. Syst.*, vol. 30, pp. 229-244, 2001.
[17] B-J Kwak, N-O Song, L. E. Miller, "Performance analysis of exponential backoff," *IEEE/ACM Trans. Networking*, pp. 343-355, April 2005.
[18] B. Tsybakov and W. Mikhailov, "Ergodicity of Slotted Aloha System, " *Probl. Inform. Transmission*, vol. 15, no. 4, pp. 73-87, 1979.
[19] R. Rao and A. Ephremides, "On the stability of interacting queues in a multiple-access system," *IEEE Trans. Inf. Theory*, pp. 918-930, 1988.
[20] V. Anantharam, "The stability region of the finite-user slotted ALOHA protocol," *IEEE Trans. Inf. Theory*, vol. 37, pp. 535-540, May 1991.
[21] W. Szpankowski, "Stability conditions for some multiqueue distributed systems: Buffered random access systems," *Adv. Appl. Prob.*, vol. 26, pp. 498-515, 1994.
[22] W. Luo and A. Ephremides, "Stability of $N$ interacting queues in random-access systems," *IEEE Trans. Inf. Theory*, vol. 45, pp. 1579-1587, Jul. 1999.
[23] J. Luo and A. Ephremides, "On the throughput, capacity, and stability regions of random multiple access," *IEEE Trans. Inf. Theory*, vol. 52, pp. 2593-2607, June 2006.
[24] J. Y. Hui and E. Arthurs, "A broadband packet switch for integrated transport," *IEEE J. Select. Areas Commun.*, vol. 5, pp. 1264-1273, 1987.
[25] M. J. Karol, M. G. Hluchyj, and S. P. Morgan, "Input versus output queueing on a space-division packet switch," *IEEE Trans. Commun.*, vol. 35, no. 12, pp. 1347-1356, Dec. 1987.
[26] J. Y. Hui, *Switching and Traffic Theory for Integrated Broadband Networks*, Kluwer Academic Publishers, 1990.
[27] D. P. Bertsekas and R. G. Gallager, *Data Networks*, Wiley.
[28] G. Bianchi, "Performance analysis of the IEEE 802.11 distributed coordination function," *IEEE J. Select. Areas Commun.*, vol. 18, no. 3, pp. 535-547, Mar. 2000.
[29] R. M. Corless, G. H. Gonnet, D. E. G. Hare, D. J. Jeffrey, and D. E. Knuth, "On the Lambert W function," *Adv. Comput. Math.*, vol. 5, pp. 329–359, 1996.
[30] R. Rom and M. Sidi, *Multiple Access Protocols: Performance and Analysis*, Springer-Verlag New York Inc., 1990.